# Surface plasmon driven scalable low-loss negative-index metamaterials at visible spectrum

**Muhammad I. Aslam**[1,2,*] and **Durdu Ö. Güney**[1]

[1] *Department of Electrical and Computer Engineering, Michigan Technological University, Houghton, Michigan 49931, USA*

[2] *Department of Electronic Engineering, NED University of Engineering and Technology, Karachi 75270, Pakistan*

* Corresponding author: maslam@mtu.edu

We demonstrate that surface plasmons of a thin metal film interacting with a periodic array of nano-structures around it can be utilized to make bulk negative index metamaterials at visible spectrum with simultaneously negative permittivity and permeability. These surface plasmon driven metamaterials have high figure of merit and can be tuned arbitrarily to operate at any wavelength in the visible spectrum and possibly at ultraviolet spectrum. We numerically demonstrate the idea by a metamaterial structure which exhibits a strong magnetic response resulting in a negative index of refraction in the green region of the electromagnetic spectrum at 536nm with a figure of merit of 3.67. We also demonstrate by simply changing the constituent material only, hence by modifying the underlying surface plasmon dispersion, that the operating wavelength of the structure can be blue-shifted to the violet region at 406nm with a figure of merit of 2.27. In contrast to the fishnet-structure based approaches for visible metamaterials, our proposed approach offers a more frequency scalable way of achieving negative index of refraction in the visible and possibly at ultraviolet wavelengths with high figure of merit.

## I. INTRODUCTION

More than three decades after the theoretical analysis of negative refractive index $(n = n' + jn'')$ [1], Pendry's work on its potential application as perfect lens[2] attracted the attention of significant number of researchers towards the design and fabrication of metamaterials in different frequency regimes of the electromagnetic spectrum. Metamaterials are engineered materials that have properties usually not available in nature or in the constituent materials. This work is focused on the metamaterials having periodic metal-dielectric arrangements with sub-wavelength unit cells. The effective permittivity $(\varepsilon = \varepsilon' + j\varepsilon'')$ and permeability $(\mu = \mu' + j\mu'')$ of these metamaterials can be controlled almost arbitrarily by varying their underlying constituents and/or the geometry[3]. The research in the area of metamaterial is motivated by the wide variety of their potential applications such as, high precision lithography[4], high resolution imaging[2, 5], invisibility cloaks[6], small antennas[7], optical analog simulators[8, 9], and quantum levitation[10].

The electromagnetic waves impinging on a metal surface produce surface waves along the metal-dielectric interface when interact with the collective oscillations of free electrons in the metal. These surface waves referred to as surface plasmon polaritons (SPP), can be exploited to achieve electric and magnetic resonators for metamaterials[11-13]. Different metamaterials are reported in the last decade to operate at different frequencies in RF and optical regimes[14] but still there is need for low-loss metamaterials at visible and ultraviolet (UV) spectrum to take advantage of metamaterial functionality to full extent. The magnetic response is generally very weak in the visible spectrum and the losses are very high, resulting in very low figure of merit (FOM), defined as $-n'/n''|_{n'=-1}$. Although the condition for negative index metamaterial (NIM) is $\varepsilon'|\mu|+\mu'|\varepsilon|<0$ [15], it is desirable to have double negative (DNG) metamaterials i.e. $\mu'<0$ and $\varepsilon'<0$ simultaneously, to achieve a high FOM[16]. So far NIMs operating in the visible spectrum have been demonstrated mostly using so-called double fishnet[17, 18] structures but to our knowledge, $n'\leq -1$ and DNG was not achieved[19-23] except for one case[23], where DNG metamaterial was demonstrated for red and near infra-red with FOM as high as 3.34.

Fortunately, there exists a different approach to get strong magnetic response utilizing the surface plasmon polaritons in the visible spectrum. We have already demonstrated that the surface plasmons on a thin metal film can interact with the nearby non-resonant structures to generate strong magnetic and electric responses in the visible spectrum[11]. In this article we demonstrate that our approach[11] can be extended to achieve DNG metamaterials having strong magnetic response and high FOM in the visible spectrum. The operating frequency is limited by the existing fabrication techniques. If allowed by the available fabrication techniques, our approach can be used to get DNG metamaterials not only in the visible but possibly in the UV regions of the electromagnetic spectrum as well. We demonstrate our idea with an example SPP driven NIM operating in the green region at 559THz (536nm) with FOM of 3.67 and investigate the underlying physical phenomena. The operating frequency (wavelength) in this article, refers to the frequency (wavelength) at which proposed NIM exhibits $n'=-1$. We also demonstrate that our metamaterial structure is very flexible and can easily be tuned/optimized for desired operating frequency. To show the scalability of proposed NIM, we present another example operating in violet region at 738THz (406nm) with FOM of 2.27. We should emphasize that our proof-of-principle structure can be fabricated with state-of-the-art nanolithography.

## II. MODEL DESCRIPTION

### A. Theoretical basis

The dispersion relation of the surface plasmons on a thin metal film surrounded by insulator shows two branches corresponding to the symmetric and antisymmetric modes depending on the current distribution at the metal surfaces (Ref. 24, Fig. 4). Due to momentum mismatch the free space photons cannot directly interact with the surface plasmons. A common approach to overcome this momentum mismatch is by using periodic metallic corrugation layers.

We choose to work on the symmetric SPP mode[11, 24] present at the surface of a gold film of thickness $t$ = 5nm. The background material is chosen to be polyimide ($\varepsilon$ = 3.5). We use periodic arrays of rectangular gold stripes on both sides of the metallic thin film. These periodic rectangular stripes serve dual purpose: First to excite the SPP modes on the metal thin film by overcoming the momentum mismatch between the surface plasmon and the incident plane wave[25, 26]. Second, the stripes interact with surface plasmons and contribute to net magnetic response[11] resulting in negative $\mu'$ which contributes to negative $n'$ in the visible.

### B. Physical geometry

First we choose SPP wavelength $\lambda_{SPP} = 80$nm as our operating point on the SPP dispersion curve of 5nm gold film embedded in polyimide. This choice is sufficiently far from the light line and also satisfies the homogeneous effective medium approximation[27, 28]. The cross-section of the unit cell of the structure is shown in Fig. 1. Periodic arrays of period $p = \lambda_{SPP}$ of rectangular gold stripes of thickness $s = 11$nm are placed above and below the thin film. The edges of the stripes are perfectly aligned with each other therefore the width of each metallic stripe is $w = p/2 = 40\text{nm}$. The coordinate axes and the polarization of normally incident field are also shown in Fig. 1. The structure has the translation invariance along the direction of incident magnetic field (z-direction in Fig. 1). The unit cell size in the propagation direction is $d = 100\text{nm}$.

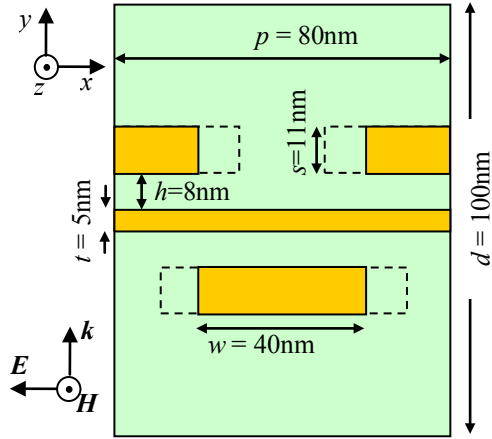

**Fig. 1.** (Color online) Geometric structure of the proposed NIM with all dimensions marked. The structure has translation invariance along the z-direction. As indicated, the **E**-field of the (normally) incident plane-wave is polarized perpendicular to the translation symmetry of the structure and **H**-field is parallel to it. Gold color in the figure represents gold described by the Drude model. The background material is polyimide having $\varepsilon = 3.5$.

Frequency domain analysis of finite element method based commercially available COMSOL Multiphysics software is used to evaluate the s-parameters corresponding to the reflection and transmission coefficients at input and output ports of the unit cell of the structure, respectively. These s-parameters are then used to evaluate the effective parameters. For numerical purposes gold is described by Drude model with the bulk plasma frequency of $f_p = 2175\text{THz}$ and the collision frequency $f_c = 6.5\text{THz}$ [29]. We use ports at the top and bottom boundaries of the computational domain shown in Fig.1. At the left and right boundaries we use periodic boundary conditions and ensure that the width of the unit cell is equal to a full-wavelength of the SPP mode. We use triangular meshes with minimum element size of 0.03nm to give at least 20 mesh elements per wavelength for all frequencies. Under these settings, a typical 2D simulation with 100 frequency samples takes about one minute on a 3GHz processor with 2GB RAM. From computational point of view, our structure has an advantage over many other metamaterials, since it does not require expensive 3D simulations owing to translational invariance in the *z*-direction.

## III. EFFECTIVE PARAMETERS AND UNDERLYING PHYSICAL PHENOMENA
### A. Effective parameters

The effective parameters ($\varepsilon$, $\mu$, *n* and *z*, where *z* is the impedance) are retrieved from the transmission and reflection coefficients using isotropic retrieval procedure[27, 28, 30]. In Fig. 2 we show the retrieved effective $\varepsilon$, $\mu$ and *n* for the proposed NIM. The structure shows a magnetic response strong enough to produce negative $\mu'$ with a minimum value reaching −8.9 where the structure also exhibits negative $\varepsilon'$, resulting in DNG metamaterial operating at 559THz. Corresponding index ($n'$) is negative from 534THz to 577THz with a minimum value of −2.3 and a FOM of 3.67. The periodicity artifacts[31] give rise to the negative imaginary parts for the retrieved effective parameters (see Fig. 2(a)). However these negative imaginary parts do not violate the passivity of the effective medium[31, 32]. The ratio of the free space wavelength to the structure size in the propagation direction ($\lambda/d$ ratio) is 5.3 which comply with the homogeneous effective medium approximation. However, if needed, this ratio may be significantly increased by removing the extra dielectric layer from top and bottom of the proposed structure (decreasing *d*). It is worth mentioning that decreasing *d* does not significantly affect the performance of the NIM.

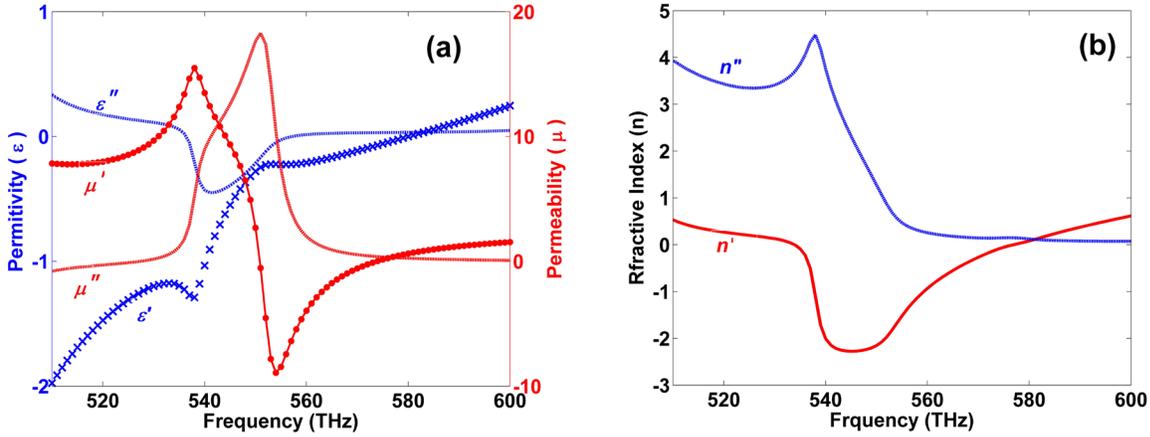

**Fig. 2**. (Color online) Retrieved effective parameters for the NIM design of Fig. 1. **(a)** Effective permittivity and effective permeability. The left and right axes show the values for $\varepsilon$ and $\mu$ respectively. Both $\varepsilon'$ and $\mu'$ are simultaneously negative with $\mu'$ having a minimum value of –8.9. **(b)** Effective refractive index. $n'$ is negative for a band of approximately 43THz (534THz to 577THz) with a minimum value of –2.3 and a FOM of 3.67 operating in the green light at 559THz.

### B. Working Principle

The array of metallic stripes overcomes the momentum mismatch between the free space and SPP modes on the metal surface and couple the light in and out of the structure. Hence, this coupling process gives rise to the extraordinary transmission (EOT) through the (otherwise opaque) metal film[33, 34]. Our simulations show that the transmittance exceeds 66% in the negative index band for the geometry shown in Fig.1.

The magnetic field profile and the induced electric currents in the metallic layers of a single unit cell of the simulated structure are shown in Fig. 3. Since we choose to work on the symmetric surface plasmon mode, currents at both metal surfaces are in the same direction and represented by single black arrows. Intuitively, the incident wave excites both the SPP mode on the thin film and the LSP mode at the metallic rectangular stripes. As evident from Fig. 3, the magnetic fields due to SPP and LSP in regions between the thin film and the rectangular stripes (indicated as ①, ② and ③), oppose each other resulting in a reduced magnetic strength in these regions, while no such cancellation occurs in the regions indicated as ④, ⑤ and ⑥. The difference in magnetic strengths at different regions of the proposed structure results in a net magnetic response capable of producing negative permeability as shown in Fig. 2. The circular loop currents on the metal surfaces contributing to magnetic moment are indicated by arrows around the regions ④, ⑤ and ⑥ in Fig. 3.

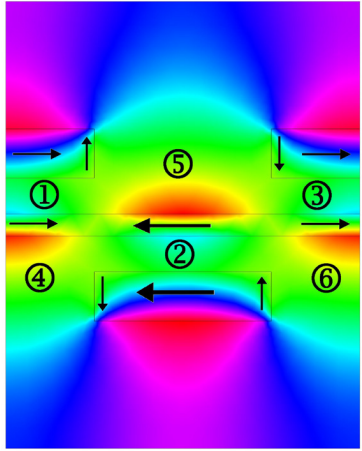

**Fig. 3**. (Color online) Surface plot shows the magnetic field (*z*-component) at the absorption peak (at 569THz) corresponding to the negative index mode of the proposed NIM. The black arrows show the current direction in the metallic layers, under normal plane-wave incidence. The surface plasmon polariton mode of the gold thin film and the localized surface plasmon mode at the rectangular stripes are clearly visible. The magnetic fields due to localized surface plasmons and thin film surface plasmons at the regions between the gold thin film and the rectangular stripes suffer destructive interference and cancel each other resulting in a net magnetic moment

strong enough to achieve negative effective permeability at the frequencies where $\varepsilon'$ is also negative. Without this destructive interference strong magnetic response is not possible.

### C. Effective index for multiple unit cells

The retrieved effective index ($n'$) for multiple unit cells in the propagation direction is shown in Fig. 4. The isotropic retrieval procedure results in multiple branches for $n'$ given by[27]

$$n' = \text{Re}\left(\frac{\cos^{-1}\left[(1-R^2+T^2)/2T\right]}{kd}\right) + \frac{2\pi m}{kd} \quad (1)$$

where $m$ is an integer and $k$ is the wavenumber. Care must be taken in choosing the correct branch especially for multilayer structure. One should choose the branches that overlap for different values of $d$ as the correct $n'$ branch[35].

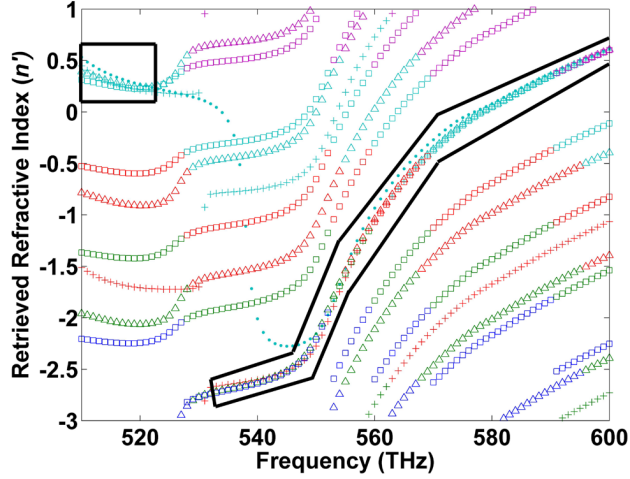

**Fig. 4**. (Color online) Retrieved real part of the refractive index for (●) one unit cell, (+) three unit cells (Δ) five unit cells, and (□) seven unit cells along the propagation direction. Branches with $m = -3$ (blue), $m = -2$ (green), $m = -1$ (red), $m = 0$ (cyan), $m = 1$ (magenta) are shown. The convergent branches corresponding to different unit cells are enclosed in a black thick line. The negative index branch for different number of unit cells converge at the region of interest.

As shown in Fig. 4, the retrieved effective $n'$ for different number of unit cells in the propagation direction converge at the region of interest (559THz). Although, the retrieved index up to seven unit cells is shown in Fig. 4, we simulated up to nineteen unit cells in the propagation direction and found the consistent behavior. Therefore the proposed NIM behaves as a bulk metamaterial[35, 36] having consistent behavior at region of interest with different number of unit cells in the propagation direction.

### D. Anti-crossing effect

Although the theoretical value of the SPP resonance frequency corresponding to $\lambda_{SPP} = 80nm$, at 491THz[24] differs substantially from the operating frequency of the proposed NIM at 559THz, the NIM is actually driven by the surface plasmons of the thin metal film similar to Ref. 11. This large difference is due to the strong coupling of the SPP mode and the intrinsic LSP modes of the rectangular stripes at 481THz (for the dimensions shown in Fig. 1). The presence of the two nearby resonant modes results in an anti-crossing behavior and thus frequency splitting in the proposed NIM[26, 37, 38]. The LSP resonance is strongly dependent on the geometry of the physical structure[39] and can be easily tuned by changing the physical dimensions ($w$, $s$, or $h$) of rectangular stripes. For instance if the width of the rectangular stripes ($w$) is increased, the LSP resonance frequency reduces and the negative index mode shifts towards the SPP resonance frequency. However, in this case the edges of the rectangular stripes will not be perfectly aligned and there will be an overlap region between rectangular stripes on both sides of the thin film (as shown by dotted lines in Fig. 1). Fig. 5 shows the observed anti-crossing behavior among two nearby resonances (SPP and LSP) in the proposed NIM as a function of the width of the rectangular stripes ($w$). Slight deviation from the theoretical anti-crossing behavior of two coupled oscillators noticeable in one of the branches is due to the presence of other resonances at higher frequencies. Properly exploiting the anti-crossing behavior in the proposed NIM, the operating frequency can be tuned in a wide range. For example having $w = 50$nm in the structure shown in Fig.1, results in DNG metamaterial to operate at 516THz (yellow light) with FOM of 3.59.

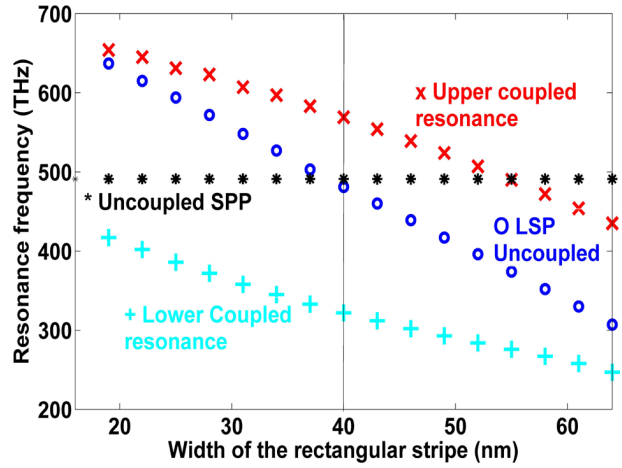

**Fig. 5**. (Color online) Observed anti-crossing behaviour between the intrinsic resonances of the rectangular stripes and the thin film surface plasmon resonances. The resonance frequencies in the graph corresponds to the absorption peaks in the respective spectra. The uncoupled LSP resonance frequency due to rectangular gold stripes is strongly dependent on the physical dimensions of the structure. This is calculated for different values of $w$ by simulating the structure shown in Fig. 1 without the thin film. The uncoupled SPP resonance frequency is independent of $w$ and therefore remains constant at its theoretical value of 491THz. The simulation of the entire structure in Fig. 1 shows two (upper and lower) coupled resonances with an observed anti-crossing and resultant frequency splitting. The vertical line indicates the operating point of our NIM at $w = 40$nm, where the negative index mode is on the upper branch.

In order to check that the proposed NIM is driven by the SPP, we analyze the convergence of the individual resonances by increasing *h*. As *h* is increased, the coupling between SPP and LSP is decreased. As a result, the frequency splitting decreases and both branches of the coupled modes (shown by + and ×) converge to intrinsic uncoupled resonances (not shown). When *h* is significantly large (>40nm), the SPP-LSP coupling becomes very weak and the coupled branch corresponding to the negative index mode converge to the theoretical surface plasmon resonance. This observation shows that the proposed NIM is indeed driven by SPP but due to the close proximity of magnetic resonance of the nearby stripes, frequency splitting occurs and the negative index mode shifts to the higher frequency as compared to its expected theoretical value.

## IV. DISCUSSION

As discussed earlier, our proposed NIM is driven by SPPs residing on the surface of a thin metal film. The operating frequency of the NIM shifts from theoretical SPP resonance frequency due to the anti-crossing behavior. The operating frequency can be changed over a wide range by properly exploiting the anti-crossing. However, having metamaterial driven by SPPs provides great flexibility in tuning the operating frequency by selecting an appropriate $\lambda_{SPP}$ from the SPP dispersion curve. In addition to this, the proposed NIM can also be tuned to any frequency in the visible spectrum by choosing appropriate constitutive materials and thin film thickness as the SPP dispersion relation depends on these parameter[24]. To demonstrate this idea, we simulate the structure in Fig. 1, with a lower index dielectric, Magnesium Fluoride ($MgF_2$; $\varepsilon = 1.9$). The physical dimensions are the same as shown in Fig. 1, except that *d* = 80nm to improve *λ*/*d* ratio. We replaced gold by silver, as it is less lossy at high frequencies[40] and from the fabrication point of view $MgF_2$ can be easily deposited on the silver[41, 42]. For simulation purposes silver is described by the Drude model with $f_p = 2180\text{THz}$ and $f_c = 13.5\text{ THz}$. This 'modified' structure shows magnetic response in the violet region of the electromagnetic spectrum to produce DNG metamaterial that operate at 738THz. The effective index ($n'$) is negative for a band of approximately 70THz (702THz to 772THz, very close to UV) with a minimum value of $-1.95$ and FOM 2.27 (see Fig. 6).

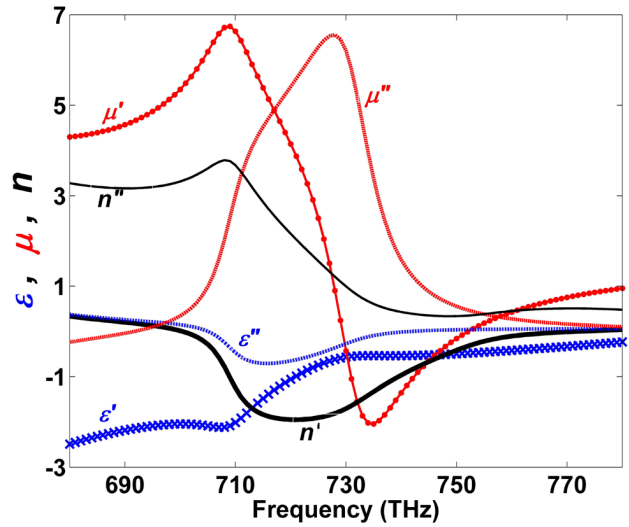

**Fig. 6**. (Color online) Retrieved effective parameters for the geometry shown in Fig. 1 with $d = 80$nm. Metal and dielectric used are silver (described using the Drude model) and Magnesium Fluoride ($\varepsilon = 1.9$) respectively. The structure is DNG metamaterial to operate at 738THz with FOM of 2.27. The corresponding $n'$ is negative for a band of 70THz (702THz to 772THz, very close to UV) with a minimum value of $-1.95$ and the $\lambda/d$ ratio is 5.04.

The reduced losses in the proposed design are due to the high-Q magnetic resonance achieved by the excitation of the SPP modes near the extended metal surface at very high frequencies based on the simple thin film dispersion relation. In contrast, low order spoof plasmon resonances, responsible from magnetic response in fishnet metamaterials cannot be easily scaled to higher frequencies and yet keeping the resonances below and close to the effective plasma frequency (i.e., low-loss region). Thus, underlying SPP dispersion of the thin metal film provides a very flexible and simple scaling mechanism for the low-loss NIMs to operate at higher frequencies. To the best of our knowledge our approach gives the highest frequency of operation so far. Our proof-of-principle structure can be fabricated with state of the art nanolithography techniques. Additionally, our proposed structure can easily be extended to get a polarization independent NIM (at normal incidence) by combining the structure shown in Fig. 1 with its orthogonal replica. We checked that such an isotropic structure is capable of producing negative index of refraction in the visible spectrum however the FOM is less than its non-isotropic counterpart.

It should be noted that a small blue shift in the plasmon resonance and a small reduction in the FOM may be possible for the fabricated structure due to the small spatial nonlocality[43, 44] expected in our structure. The effect should become more prominent at higher frequencies due to smaller particle sizes and inter-particle distances.

It should be also noted that the retrieval procedure normally neglects the spatial dispersion in the simulated metamaterial structure. However, this produces periodicity artifacts such as negative imaginary parts in the effective permittivity and permeability as rigorously studied in Ref. 31. Observed negative imaginary parts in the regions, where we obtain real negative effective permittivity, show that not only Drude-like electric response of the thin metal film but also multiple scattering (i.e., periodicity) has contribution to the effective permittivity. In fact, the contribution due to the multiple scattering should not be surprising, because we observe two EOT windows corresponding to the two resonant surface plasmon coupled modes separated by a gap as a result of anti-crossing in Fig. 5. Nevertheless, Fig. 4 clearly demonstrates that our structure possesses a large and length-independent (i.e., bulk) negative index of refraction.

## V. CONCLUSION

Utilizing the interaction of SPP of a thin metal film with a periodic array of rectangular stripes around it we demonstrate DNG-NIM with high FOM to operate at visible spectrum and possibly scalable to UV wavelengths. We also presented the analysis of underlying physical phenomena to show that the proposed structure is driven by thin film SPP and the operating frequency differs from theoretical SPP resonance due to anti-crossing. By controlling the SPP dispersion relation or by exploiting the anti-crossing behavior, the operating frequency of the proposed NIM can easily be tuned to any desired value, limited by the current fabrication technology. We presented theoretical results and retrieved effective parameters of two NIMs operating at 559THz (536nm) with FOM 3.67 and 738THz (406nm) with FOM 2.27. The proposed structure can be extended to obtain polarization-independent isotropic NIMs. In contrast to the fishnet based approach to metamaterials, our approach provides more scalable NIMs to operate in the visible spectrum and possibly at UV with substantially large FOM.

## ACKNOWLEDGMENT

We would like to thank Daw Don Cheam at Michigan Tech for discussion on possible fabrication of our proposed structure and and Thomas Koschny at Ames national laboratory for valuable discussion on impact of periodicity. M. I. A. would like to thank NED University and Daniel R. Fuhrmann at Michigan Tech for their corresponding financial support for his PhD research.